\documentclass[pdflatex,sn-mathphys-num]{sn-jnl}

\usepackage{graphicx}
\usepackage{multirow}
\usepackage{amsmath,amssymb,amsfonts}
\usepackage{amsthm}
\usepackage{mathrsfs}
\usepackage[title]{appendix}
\usepackage{xcolor}
\usepackage{textcomp}
\usepackage{manyfoot}
\usepackage{booktabs}
\usepackage{algorithm}
\usepackage{algorithmicx}
\usepackage{algpseudocode}
\usepackage{listings}
\usepackage{upgreek}
\usepackage{anyfontsize}
\usepackage{titlesec}  
\usepackage{enumitem}
\usepackage{chngcntr}
\counterwithout{table}{section}

\titleformat{\section}{\fontsize{12}{12}\bfseries}{\thesection}{0.5em}{}
\titleformat{\subsection}{\fontsize{12}{12}\bfseries}{\thesubsection}{0.5em}{}

\newcommand{\vct}[1]{\boldsymbol{#1}}
\newcommand{\dct}[1]{\mathbf{#1}}
\newcommand{\eb}[1]{\dct{e}_{#1}}
\newcommand{\db}[1]{\dct{d}_{#1}}

\definecolor{orangebeta}{HTML}{FF7514}

\theoremstyle{thmstyleone}

\theoremstyle{thmstyletwo}

\theoremstyle{thmstylethree}

\newcommand{\zero}[1]{#1^{(0)}}
\newcommand{\uno}[1]{#1^{(1)}}

\usepackage[normalem]{ulem}
\begin{document}

\title[]{
\centering A three-dimensional morphoelastic model \\for self-oscillations in polyelectrolyte hydrogel filaments
}

\author[2]{\fnm{A.\,S.} \sur{Boiardi}}\equalcont{These authors contributed equally to this work}

\author[1]{\fnm{R.} \sur{Marchello}}\equalcont{These authors contributed equally to this work}

\author[1]{\fnm{P.\,M.} \sur{Santucci}}

\author[1]{\fnm{D.} \sur{Riccobelli}}

\author*[1]{\fnm{G.} \sur{Noselli}}
\email{giovanni.noselli@sissa.it}

\affil[1]{\centering
\orgname{\small SISSA\,--\,International School for Advanced Studies}, \orgdiv{Mathematics Area, mathLab},\\ \orgaddress{Via Bonomea 265, \city{Trieste}, \postcode{34136}, \country{Italy}}}

\affil[2]{\centering
\orgname{\small Okinawa Institute of Science and Technology}, \orgdiv{Mechanics and Materials Unit}, \\ \orgaddress{1919-1 Tancha, \city{Onna}, \postcode{Okinawa 904-0495}, \country{Japan}}}

\abstract{
We introduce a three-dimensional model for polyelectrolyte hydrogel filaments operating in a fluid environment under an electric field. 
The formulation builds on a morphoelastic framework for inextensible and unshearable rods, such that the filament’s activity is encoded in electric-field-induced spontaneous curvatures, while hydrodynamic interactions are captured via a local approximation of Stokes flows. 
We employ this framework to investigate the prototypical case of a filament with elliptical cross-section clamped at its base. 
Under a constant and uniform electric field aligned with its axis, the filament undergoes flutter instability beyond a critical field strength, as revealed by a linear stability analysis. 
Depending on the model parameters, the instability is characterized by either two- or three-dimensional self-sustained oscillations. 
We further examine this behaviour through numerical simulations in the post-critical regime, showing that flutter may develop into large amplitude planar oscillations or more complex three-dimensional motions, through a secondary bifurcation. 
Although the study represents a first step towards extending state-of-the-art models for polyelectrolyte hydrogel filaments to three dimensions, the richness of the resulting dynamics achievable under time-independent forcing underscores the potential of the proposed actuation mechanism for the design of biomimetic cilia and soft robotic systems.
}

\keywords{
Active filaments,
Flutter instability,
Morphoelasticity, 
Polyelectrolyte hydrogels,
Soft robotics}

\maketitle

\section{Introduction}
\label{INTRO}

Soft active materials, such as hydrogels and liquid crystal elastomers, have attracted considerable attention over the past decades due to their ability to undergo controlled deformation in response to environmental stimuli. This remarkable property has motivated significant research efforts aimed at developing continuum theories capable of describing the multi-physics behaviour of this class of materials~\cite{warner_terentjev_2003, suo_2008,suo_2010, doi_2013}. In parallel, a substantial body of literature has emerged focusing on the potential applications of active materials across a wide range of engineering fields~\cite{klein_efrati_sharon_2007,sharon_2009, mahadevan_2016, damioli_2022, biggins_2025}.

In soft robotics, active materials find application in the development of biomimetic cilia and flagella, as their intrinsic responsiveness to external stimuli replaces the function of the molecular motors responsible for actuation in the biological counterpart~\cite{nicastro_2018}.
In nature, cilia and flagella are capable of generating coordinated beating patterns that enable fluid transport, particle manipulation, and locomotion at low Reynolds numbers.
Replicating these functionalities in artificial systems has therefore attracted increasing interest for applications in microfluidics, lab-on-chip technologies~\cite{liu20263d}, and bioinspired robotic devices~\cite{Dong2020}. 
A major challenge in this field remains the achievement of untethered control over such biomimetic systems to enable programmable and dynamic shape transformations. These are achievable through the application of complex external controls, which are modulated either in space or in time.
While effective, these strategies underscore the need for simpler, and hence more scalable, modes of control. 

The ability of biological systems to self-excite plays a fundamental role in both the origin and the maintenance of life~\cite{Noble_CardiacActionPacemaker_1960}. In this regard, dynamical instabilities have been recognized as the key mechanism underlying several self-sustained oscillatory behaviours observed in living systems, spanning multiple scales, from the periodic beating of flagella~\cite{JulicherProst_CooperativeMolecularMotors_1995,BaylyDutcher_SteadyDyneinForces_2016} to rhythmic processes in growing plants~\cite{agostinelli_2020}.
In artificial systems, feedback loops have also been identified as a means to achieve self-oscillatory responses by harnessing mechanical instabilities, as exemplified by the Quincke rotation~\cite{Stone2021} and the self-shadowing mechanisms~\cite{Zhao2019,agostinelli_2026}.

Polyelectrolyte hydrogels~\cite{suo_2010} constitute a class of soft active materials capable of undergoing large deformations when subjected to electric stimuli. 
In a series of works~\cite{cicconofri_2023,boiardi_marchello_2024,boiardi_2024}, some of the authors have demonstrated that a polyelectrolyte hydrogel filament can exhibit self-sustained oscillations originating from flutter instability~\cite{Bigoni_Flutter_2023} in response to a constant and uniform electric field.
Owing to the non-reciprocal nature of the resulting oscillatory motion, such an actuation strategy can be harnessed for locomotion~\cite{boiardi_2024} and transport~\cite{cicconofri_2023,boiardi_marchello_2024} at low Reynolds number, where the celebrated scallop theorem by Purcell~\cite{purcell_1977} precludes locomotion by time-reversible deformation cycles.
In this context, flutter instability, rather than being a source of structural failure, enables the embodiment of mechanical intelligence~\cite{KacprzykPedrycz_SpringerHandbookComputational_2015} in robotic systems, such that low-level control tasks are offloaded to the intrinsic response of the constituent material to the electric stimulus. 

These studies were restricted to hydrogel filaments in planar motion, which were also realized experimentally by employing ribbon-like geometries. 
This simplified configuration allowed for the exploration of the flutter-based actuation strategy within accessible experimental setups and tractable mathematical models.
On the other hand, constraining the dynamics to planar motions significantly limits the range of achievable deformations and, consequently, the richness of the resulting actuation mechanisms.
Relaxing this constraint is therefore a necessary step toward developing more physically relevant models and unlocking novel design principles for hydrogel-based biomimetic actuators.

To this end, in this work we propose a three-dimensional morphoelastic model for polyelectrolyte hydrogel filaments with elliptical cross-section.
This model consistently extends previous results by the authors, allowing us to explore a wider range of configurations and motions, in addition to reproducing established findings.
While here we focus on self-oscillatory dynamics of a single filament clamped at the base for comparison with~\cite{cicconofri_2023}, future work will be devoted to free-swimming filaments in space and other configurations.

The manuscript is organized as follows.
In Section~\ref{MODEL} we develop the mathematical model based on Kirchhoff rod theory with natural strains, which couple the mechanical response of the filament to the electric stimulus.
In Section~\ref{STAB} we study the linear stability analysis of a filament aligned with a constant external electric field and predict the onset of flutter while varying the model parameters. The nonlinear dynamics of the system is analysed via numerical simulations in Section~\ref{EXP}. Here particular attention is devoted to identifying the range of model parameters in which two- and three-dimensional motions coexist and to comparing them in terms of energetics.
Finally, we propose some perspectives and concluding remarks in Section~\ref{CONC}.

\section{A model for active filaments deforming in space}
\label{MODEL}
We introduce a mathematical model to describe the active response of polyelectrolyte hydrogel filaments to an electric stimulus, while accounting for their hydrodynamic interactions with the surrounding viscous fluid. As detailed later, we adopt a morphoelastic formulation of Kirchhoff rod theory~\cite{antman_2005,goriely_2017}, whereby the filament's activity is encoded via spontaneous strains. In the formulation, these couple the filament's mechanical response to the environmental electric stimulus.

\subsection{Kinematics}
\label{MODEL:kin}

\begin{figure}
    \centering
    \includegraphics[width=\textwidth]{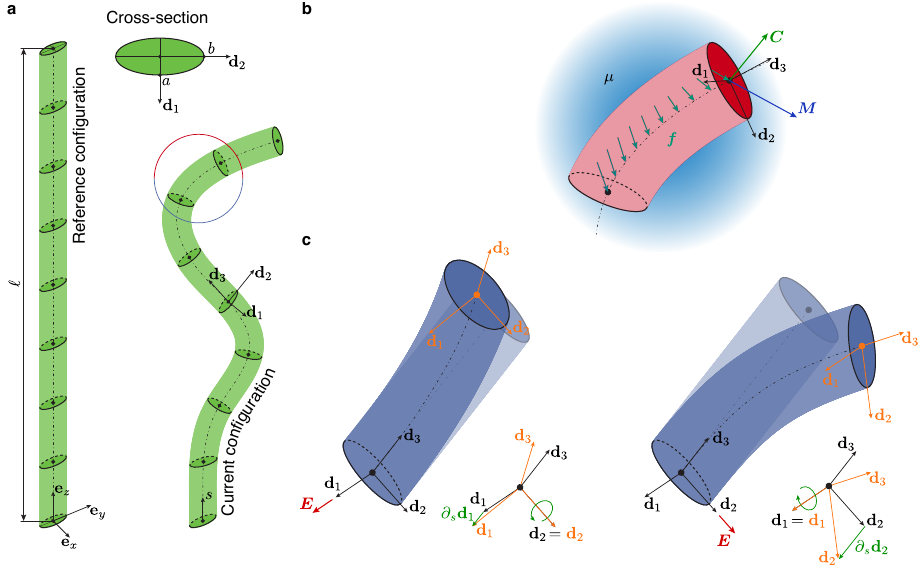}
    \caption{
    Reference and current configurations of an inextensible and unshearable rod of length $\ell$ and elliptic cross-section with semi-axes $a$ and $b$.
    In (a) notice the triplet of orthonormal directors $\dct{d}_i$, $i=\{1,2,3\}$. The rod is acted upon by drag forces per unit length $\vct{f}$, arising from its hydrodynamic interaction with the surrounding fluid of dynamic viscosity $\mu$. The vectors $\vct{C}$ and $\vct{M}$ in (b) denote the internal contact force and moment, respectively. The application of an electric field $\vct{E}$ in the cross-sectional plane causes the active bending of the hydrogel rod, as shown in (c).}
    \label{fig:kinematics}
\end{figure}

Consider a morphoelastic rod of length $\ell$ and uniform cross-section that is straight in the reference configuration.
To describe its current configuration, we introduce the mapping $(s,t) \mapsto \vct r(s,t)$ for the \emph{centreline} and a triplet of orthonormal \emph{directors} $\dct{d}_i(s,t)$, $i=\{1,2,3\}$, such that $\dct{d}_1(s,t)$ and $\dct{d}_2(s,t)$ identify the principal axes in the cross-sectional plane and $\dct{d}_3(s,t) = \dct{d}_1(s,t) \times \dct{d}_2(s,t)$, see Fig.~\ref{fig:kinematics}a. Here $s\in[0,\ell]$ denotes the arc-length parameter and $t\in [0,\infty)$ time.
We assume that the rod's length is much greater than the relevant cross-sectional dimensions, and model it as both unshearable and inextensible.
By these kinematic constraints $\partial_s \vct r (s,t) = \dct{d}_3(s,t)$, while the velocity is introduced as $\vct{v}(s,t) = \partial_t \vct{r}(s,t)$.

Since the directors form a right-handed orthonormal basis, there exists a vector field $\vct{u}(s,t)$, called \emph{Darboux vector}, such that
\begin{equation}
    \label{eq:vectors_uw}
    \partial_s \dct{d}_i = \vct{u}\times \dct{d}_i,  
    \qquad i=1,2,3,
\end{equation}
which encodes \emph{bending} and \emph{twist} of the rod; in particular, expressing $\vct{u} = u_i\db{i}$ in the directors' basis (summation over repeated indices is implied), the components $u_1$ and $u_2$ measure bending of the rod about the respective principal axes in the cross-sectional plane, whereas $u_3$ measures twist of the rod about its axis. 
Such strain measures will be later exploited to introduce constitutive assumptions suitable for the present morphoelastic theory of active filaments based on polyelectrolyte hydrogels.

Given the Darboux vector field $\vct{u}$, equation~\eqref{eq:vectors_uw} provides the orientation of the cross-sections of the filament along its centreline by solving a system of linear ODEs that is supplemented by the boundary conditions imposed at $s=0$. 
We note that the structure of~\eqref{eq:vectors_uw} ensures that the orthonormality of the directors is preserved at all points along the rod and at all times.
Indeed,
\begin{equation}
    \partial_s\left(\db{i}\cdot\db{j}\right) = \partial_s\db{i}\cdot\db{j} + \db{i}\cdot\partial_s\db{j} = \left(\vct{u}\times \db{i}\right)\cdot\db{j} + \db{i}\cdot\left(\vct{u}\times \db{j}\right) = 0,
\end{equation}
so that, if $\db{i}\cdot\db{j} = \delta_{ij}$ at $s=0$, then $\db{i}\cdot\db{j} = \delta_{ij}$ for all $s\in[0,\ell]$.

\subsection{Balance equations}
\label{MODEL:balance_eqs}

We neglect inertia in setting the mechanical problem, so that the balance equations of linear and angular momentum read
\begin{equation}
    \label{eq:blm}
    \partial_s\vct{C} + \vct{f} = \vct{0}, \qquad \partial_s \vct{M} + \partial_s \vct{r} \times \vct{C} = \vct{0}, 
\end{equation}
where $\vct{C}(s,t)$ and $\vct{M}(s,t)$ are the internal contact force and moment, whereas $\vct{f}(s,t)$ is the external force per unit length~\cite{antman_2005}, see Fig.~\ref{fig:kinematics}b.
Under inextensibility and unshearability assumptions, the internal contact force acts as a reaction enforcing such kinematic constraints. In contrast, the internal moment is determined by the constitutive equations detailed in the next section according to morphoelasticity. 
As for the external force per unit length, this encodes the hydrodynamic interactions of the rod with a surrounding fluid of dynamic viscosity $\mu$. Assuming low Reynolds number hydrodynamics, we adopt the local approximation of Resistive Force Theory~\cite{lauga2009hydrodynamics}, such that the viscous force is proportional to the local velocity of the rod via suitable resistive coefficients $\mu_1,\mu_2,\mu_3$, that is
\begin{equation}
    \label{eq:visc}
    \vct{f} = - \left[\mu_1(\db{1}\otimes\db{1}) +  \mu_2 (\db{2}\otimes\db{2}) + \mu_3(\db{3}\otimes\db{3}) \right] \vct{v} .
\end{equation}

The mechanical problem governed by the balance equations~\eqref{eq:blm} is complemented by suitable initial and boundary conditions. As for the latter,
here we consider the prototypical case of an active rod clamped at its base and free at the other extremity, such that the boundary conditions are of both kinematic and dynamic nature 
\begin{equation}
    \label{eq:boundary_clamped}
    \vct{r}(0,t) = \vct{0}, \quad \dct{d}_1(0,t) = \eb{x}, \quad \dct{d}_2(0,t) = \eb{y}, \quad \vct{M}(\ell,t) = \vct{0}, \quad \vct{C}(\ell, t) = \vct{0} , \quad \forall\,t\in[0,\infty) .
\end{equation}
We note that this boundary value problem is of particular relevance for the study of synthetic active cilia deforming in space, which is the main subject of the present work.

\subsection{Morphoelastic constitutive relations}
\label{MODEL:morph}

We adopt morphoelasticity theory to model the active response of the hydrogel rod to directed electric fields~\cite{goriely_2017}. In particular, we introduce the \emph{spontaneous strain vector} $\vct{\kappa}(s,t)$ and assume that the internal moment $\vct{M}(s,t)$ is proportional to the difference between the \emph{visible} and the \emph{spontaneous} strains  
\begin{equation}
    \label{eq:moment_morphoelastic}
    \vct{M} = \mathbb{B}\left(\vct{u} - \vct{\kappa} \right),
\end{equation}
where $\mathbb{B} = B_1(\db{1}\otimes\db{1}) + B_2(\db{2}\otimes\db{2}) + B_3(\db{3}\otimes\db{3})$ is the constitutive tensor encoding the bending stiffnesses about the principal axes in the cross-sectional plane, $B_1$ and $B_2$, and the torsional stiffness about the rod's axis, $B_3$.
In passing, we remark that this linear constitutive relation is appropriate for the large-deformation/small-strain regime considered in this work.

Polyelectrolyte hydrogels are known to respond to electric fields by the forced migration of mobile ionic species in the cross-sectional plane and the consequent gradient in their osmotic pressure~\cite{suo_2008,suo_2010}. 
This forced diffusive process is such that a hydrogel rod progressively bends, in the direction of the electric field, to attain a \emph{target} spontaneous curvature in a \emph{characteristic time}. 
We model such active process by extending an approach successfully exploited in the context of hydrogel rods undergoing planar motions~\cite{cicconofri_2023, boiardi_2024, boiardi_marchello_2024}. 
In doing so, we restrict attention to spontaneous curvatures, since, by the underlying forced migration process, spontaneous twist is not expected to arise from coupling with the electric field. 
Hence, we take the following form for the spontaneous strain
\begin{equation}
    \label{eq:spont_strain}
    \vct{\kappa} = \kappa_1 \db{1} + \kappa_2 \db{2} ,
\end{equation}
and assume the  spontaneous curvatures $\kappa_1$ and $\kappa_2$ to be determined by the components $E_2=\vct{E} \cdot \db{2}$ and $E_1 = \vct{E} \cdot \db{1}$ of the electric field, respectively. 
With the purpose of introducing suitable phenomenological \emph{evolution laws} for the spontaneous curvatures based on experimental evidence reported in~\cite{cicconofri_2023,boiardi_2024}, we illustrate step-by-step how the rod deforms in response to the electric stimulus.
We discuss separately the cases in which either $E_1$ or $E_2$ are null and consider the equilibrium state, so that $\kappa_1 = u_1$ and $\kappa_2 = u_2$. 
When $E_1 \neq 0$, bending takes place about the $\db{2}$-axis by the evolution of $\kappa_2$, see Fig.~\ref{fig:kinematics}c (left). 
Motivated by experimental observations, we require active bending to occur in the direction of the electric field, so that $\partial_s \db{1} \cdot \db{3} < 0$ for $E_1 > 0$. Since $\partial_s \db{1} = - \kappa_2 \db{3}$, at equilibrium $\kappa_2$ has the sign of $E_1$, and hence its evolution law reads
\begin{equation}
    \label{eq:evol_1}
    \tau_2 \partial_t \kappa_2 + \kappa_2 = \kappa_2^\star (\vct{E} \cdot \db{1}) ,
\end{equation}
where $\tau_2$ is the characteristic time of the active reconfiguration around $\db{2}$ and $\kappa_2^\star>0$ is a parameter encoding the sensitivity of the rod to the applied field, namely, the target curvature achieved for an electric field of unit magnitude along $\db{1}$. 
When, instead, $E_2 \neq 0$, bending is about the $\db{1}$-axis by the evolution of $\kappa_1$, see Fig.~\ref{fig:kinematics}c (right). 
As for the previous case, we require bending to occur in the direction of the electric field, so that $\partial_s \db{2} \cdot \db{3} < 0$ for $E_2 > 0$. 
In this case, though, $\partial_s \db{2} = \kappa_1 \db{3}$, so that at equilibrium the sign of $\kappa_1$ is opposite to that of $E_2$ and its evolution law  reads
\begin{equation}
    \label{eq:evol_2}
    \tau_1\partial_t \kappa_1 + \kappa_1 = -\kappa_1^\star (\vct{E} \cdot \db{2}) ,
\end{equation}
where again $\tau_1$ is the characteristic time and the parameter $\kappa_1^\star > 0 $ encodes the sensitivity of the rod to the applied field, in this case for an electric field of unit magnitude along $\db{2}$.
While the proposed model may describe time-varying electric stimuli, here we consider the case of a constant field, so that $\vct{E} = E_x \eb{x} + E_y \eb{y} + E_z \eb{z}$. Although the electric field is constant, its components along the directors in the cross-sectional plane may vary in time according to the configuration of the rod, as captured by the forcing terms on the right-hand side of equations~\eqref{eq:evol_1}-\eqref{eq:evol_2}.

The mechanical response of  polyelectrolyte hydrogel filaments presented here refers to polyacrylamide-co-sodium acrylate hydrogels employed for the experiments in~\cite{cicconofri_2023,boiardi_2024}.
In particular, the direction of spontaneous bending and, hence, the sign of the target curvatures as discussed above reflects the anionic nature of the specific hydrogel considered. 
We believe that, in its simplicity, this modelling framework can be adapted to other polyelectrolyte hydrogels and even to a wider class of active materials, such as liquid crystal elastomers~\cite{Korner_NonlinearBeamModel_2020,agostinelli_2026}.

\subsection{Balance equations in local components}
\label{MODEL:comp}

We now derive a representation of the governing equations in components with respect to the local directors' basis. 
We begin by writing $\vct{C}(s,t) = C_i(s,t) \,\dct{d}_i(s,t)$ for the internal contact force, so that 
\begin{equation}
    \label{eq:dC_components}
    \partial_s \vct{C} 
    = \partial_s C_i \,\db{i} + C_i\, \partial_s\db{i} 
    = \partial_s C_i\, \db{i} + C_i(\vct{u} \times \db{i})
    = \partial_s C_i \,\db{i} + C_i\, u_j \epsilon_{jik} \,\db{k} ,
\end{equation}
where $\epsilon_{jik}$ is the Levi-Civita symbol. 
Balance of linear momentum~\eqref{eq:blm}$_1$ is then expressed in local components as
\begin{equation}
    \label{eq:blm_components}
    \left\{
        \begin{aligned}
            \partial_s C_1 - u_3 C_2 + u_2 C_3 + f_1 = 0 ,\\
            \partial_s C_2 + u_3 C_1 - u_1 C_3 + f_2= 0  ,\\
            \partial_s C_3 - u_2 C_1 + u_1 C_2 + f_3= 0 ,
        \end{aligned}
    \right.
\end{equation}
where $f_i = -\mu_i v_i$, with $v_i = \vct{v}\cdot\dct{d}_i$ for $i=\{1,2,3\}$. 
We next write $\vct{M}(s,t) = M_i(s,t) \db{i}(s,t)$, so that
\begin{equation}
    \partial_s \vct{M} = \partial_s M_i \,\db{i} + M_i \,\partial_s \db{i}
    = \partial_s M_i \,\db{i} + M_i(\vct{u} \times \db{i})
    = \partial_s M_i \,\db{i} + M_i \,u_j \epsilon_{jik} \,\db{k}.
\end{equation}
Upon recalling that $\partial_s \vct{r} = \db{3}$, we also have
\begin{equation}
    \partial_s \vct{r} \times \vct{C} = C_1 \db{2} - C_2 \db{1},
\end{equation}
so that the balance of angular momentum \eqref{eq:blm}$_2$ in local components reads
\begin{equation}
    \label{eq:bam_components}
    \left\{
        \begin{aligned}
            &\partial_s M_1 - u_3 M_2 + u_2 M_3  - C_2 = 0, \\
            &\partial_s M_2 + u_3 M_1 - u_1 M_3 + C_1 = 0, \\
            &\partial_s M_3 - u_2 M_1 + u_1 M_2  = 0 .
        \end{aligned}
    \right.
\end{equation}
In the following, we proceed by expressing the governing equations in dimensionless form.
\subsection{Dimensionless form of the governing equations}
\label{MODEL:nondim}

We assume the cross-section of the hydrogel filament to be elliptic, with semi-axes $a, b \ll \ell$ oriented along $\dct{d}_1$ and $\dct{d}_2$, respectively, as sketched in Fig.~\ref{fig:kinematics}a. Then, by adopting the seminal results reported in~\cite{Batchelor_SlenderbodyTheoryParticles_1970}, the resistive coefficients in equation~\eqref{eq:visc} take the following explicit form
\begin{equation}
\label{eq:res_coef}
    \mu_1 = \frac{4\pi\mu}{\log\frac{2\ell}{a+b} + \frac{a}{a+b}} , \qquad \mu_2 = \frac{4\pi\mu}{\log\frac{2\ell}{a+b} + \frac{b}{a+b}}, \qquad \mu_3 = \frac{2\pi\mu}{\log\frac{2\ell}{a+b}-\frac{1}{2}} .
\end{equation}
Denoting by $Y$ and $\nu$ the Young's modulus and the Poisson's ratio, respectively, the bending and the torsional stiffnesses read
\begin{equation}
    B_1 = Y J_1, \qquad B_2 = Y J_2, \qquad B_3 = G J_3 ,
\end{equation}
where $G = Y/[2(1 + \nu)]$ is the shear modulus and 
\begin{equation}
    J_1 = \frac{\pi a b^3}{4}, \qquad J_2 = \frac{\pi a^3 b}{4}, \qquad J_3 = \frac{\pi a^3 b^3}{a^2 + b^2} 
\end{equation}
are the second moments of inertia and the torsional constant. We hence rewrite~\eqref{eq:moment_morphoelastic} in component form, obtaining
\begin{equation}
    \label{eq:const_comp}
    M_1 = B_1(u_1 - \kappa_1), \qquad M_2 = B_2(u_2 - \kappa_2), \qquad M_3 = B_3 u_3, 
\end{equation}
in which we accounted for the fact that coupling with the electric field does not induce active twist.

We now recast the governing equations in dimensionless form. To this end, we identify $\ell$ as the characteristic length-scale and $\tau_1$ as the characteristic time-scale, so that $\bar{s} = s\,\ell^{-1}\!\in [0,1]$ and $\bar{t} = t\,\tau_1^{-1}\!\in [0,\infty)$ denote the dimensionless arc-length and time, respectively. 
Furthermore, we normalize forces by $Y\,\ell^2$, moments by $Y\,\ell^3$, velocities by $\ell\,\tau_1^{-1}$, and strains by $\ell^{-1}$. With a slight abuse of notation, hereafter we denote dimensionless quantities with the same symbols of their dimensional counterparts and omit overbars for ease of notation.

To proceed, we introduce the dimensionless quantities $\lambda = \ell / b$ and $\eta = a / b$, and recast the resistive coefficients~\eqref{eq:res_coef} in the form $\mu_i = \mu \, \zeta_i$, where 
\begin{equation}
\label{eq:res_coef_nondim}
    \zeta_1 = \frac{4\pi}{\log\frac{2 \lambda}{1 + \eta} + \frac{\eta}{1 + \eta}} , \qquad
    \zeta_2 = \frac{4\pi}{\log\frac{2 \lambda}{1 + \eta} + \frac{1}{1 + \eta}}, \qquad 
    \zeta_3 = \frac{2\pi}{\log\frac{2 \lambda}{1 + \eta} - \frac{1}{2}} .
\end{equation}
The balance of linear momentum~\eqref{eq:blm_components} becomes
\begin{equation}
    \label{eq:blm_components_nondim}
    \left\{
        \begin{aligned}
            \partial_s C_1 - u_3 C_2 + u_2 C_3 - \xi\, \zeta_1 v_1 = 0,\\
            \partial_s C_2 + u_3 C_1 - u_1 C_3 - \xi\, \zeta_2 v_2= 0,\\
            \partial_s C_3 - u_2 C_1 + u_1 C_2 - \xi\, \zeta_3 v_3= 0,
        \end{aligned}
    \right.
\end{equation}
where the parameter $\xi = \mu / (Y \tau_1)$ can be interpreted as the ratio between a passive relaxation time of the filament in the viscous environment, $\mu/Y$, and the characteristic time of active bending, $\tau_1$.
As for the balance of angular momentum~\eqref{eq:bam_components}, this becomes
\begin{equation}
    \label{eq:bam_components_nondim}
    \left\{
        \begin{aligned}
            & \beta_1 \partial_s (u_1-\kappa_1) - \beta_2 u_3  (u_2-\kappa_2) + \beta_3 u_2  u_3 - C_2 = 0, \\
            & \beta_2 \partial_s (u_2-\kappa_2) + \beta_1 u_3 (u_1-\kappa_1) - \beta_3 u_1  u_3 + C_1 = 0, \\
            & \beta_3  \partial_s u_3 - \beta_1 u_2 (u_1-\kappa_1) + \beta_2 u_1  (u_2-\kappa_2) = 0 ,
        \end{aligned}
    \right.
\end{equation}
where we have accounted for the constitutive equations~\eqref{eq:const_comp} and introduced
\begin{equation}
\label{eq:bend_stiff_nondim}
    \beta_1 = \frac{\pi \, \eta}{4 \, \lambda^4} , \qquad
    \beta_2 = \frac{\pi \, \eta^3}{4 \, \lambda^4} , \qquad 
    \beta_3 = \frac{\pi \, \eta^3}{2 \, \lambda^4 \,  (1 + \eta^2) \, (1 + \nu)} ,
\end{equation}
as the dimensionless bending and torsional stiffnesses.

Regarding the evolution laws~\eqref{eq:evol_1} and~\eqref{eq:evol_2} for the spontaneous curvatures, these become
\begin{equation}
    \label{eq:evol_nondim_alt}
    \partial_t \kappa_1 + \kappa_1 = -\vct{\chi}_1 \cdot \db{2} ,\qquad
    \tau \,\partial_t \kappa_2 + \kappa_2 = \vct{\chi}_2\cdot \db{1} ,
\end{equation}
where 
\begin{equation}
    \tau = \frac{\tau_2}{\tau_1} , \qquad \vct{\chi}_1 = \ell \,\kappa_1^\star  \, \vct{E}, \qquad 
    \vct{\chi}_2 = \ell\,\kappa_2^\star \, \vct{E}.
\end{equation}
Notice that the parameters $\vct{\chi}_1$ and $\vct{\chi}_2$ encode both the magnitude and the direction of the external stimulus, as well as the filament's sensitivities to such stimulus.

To proceed, we recall that the active response of polyelectrolyte hydrogel filaments to electric fields is governed by the forced transport of mobile ions in the cross-sectional plane and the resulting gradient in the osmotic pressure they exert. Since diffusivity has dimensions of length squared divided by time, guided by dimensional analysis we set $\tau_1 \propto b^2$ and $\tau_2 \propto a^2$, having identified $b$ and $a$ as the relevant length-scales for the kinetic processes underlying active bending about $\db{1}$ and $\db{2}$, respectively. Likewise, since curvature has dimensions of inverse length, we set $\kappa_1^\star \propto 1/b$ and $\kappa_2^\star \propto 1/a$. Here the underlying modelling assumption is that the maximum active strain in the cross-sectional plane is thickness-independent for a given electric stimulus. According to these scalings, $\tau_1/\tau_2 = b^2/a^2$ and $\kappa_1^\star/\kappa_2^\star = a/b$, so that $\tau = \eta^2$ and $\vct{\chi}_2 = \vct{\chi}_1/\eta$.
Hence the evolution equations~\eqref{eq:evol_nondim_alt} can be recast as
\begin{equation}
    \label{eq:evol_nondim_alt2}
    \partial_t \kappa_1 + \kappa_1 = -\vct{\chi}_1 \cdot \db{2} ,\qquad
    \partial_t \kappa_2 +  \eta^{-2} \kappa_2 = \eta^{-3} (\vct{\chi}_1 \cdot \db{1}),
\end{equation}
and the set of independent dimensionless parameters reduces to $\{\nu, \lambda, \eta, \xi, \vct{\chi}_1\}$.
To facilitate reference, we report in Table~\ref{tab:adm_par} the independent dimensionless parameters of the model, together with their explicit expressions and physical interpretations.
\begin{table*}[t]
    \centering
    \begin{tabular}{ccc}
    ~~~~~~Model parameter~~~~~~ & ~~~~~~Parameter expression~~~~~~ & ~~~~~~Parameter description~~~~~~\\[1mm]
    \toprule
    $\lambda$   &       $\ell/b$            &   slenderness of the filament \\[1.5mm]
    $\eta$      &       $a/b$               &   cross-sectional aspect ratio \\[1.5mm]
    $\xi$       &       $\mu/(Y \tau_1)$    &   passive/active time ratio \\[1.5mm]
    $\nu$       &       -    &   material Poisson's ratio \\[1.5mm]
    $\vct{\chi}_1$       &       $\ell \,\kappa_1^\star\,\vct{E}$    &   external driving stimulus \\
    \bottomrule
    \end{tabular}
    \medskip
    \caption{Independent model parameters as introduced by recasting the governing equations in dimensionless form.}
    \label{tab:adm_par}
\end{table*}
\section{Stability analysis of a filament aligned with the electric field}
\label{STAB}

Let us consider the case in which the electric field is applied along the filament's axis in the reference configuration, such that $\vct{E}=E_z\eb{z}$, see Fig.~\ref{fig:kinematics}a.
Under this condition, the straight configuration is of equilibrium. We denote the fields corresponding to this solution of the governing equations by the superscript \lq ${(0)}$', so that
\begin{equation}
    \vct{r}^{(0)} = s \eb{z}, ~~~\db{1}^{(0)} = \eb{x}, ~~~ \db{2}^{(0)} = \eb{y}, ~~~ \db{3}^{(0)} = \eb{z}, ~~~
    \vct{C}^{(0)} = \vct{0}, ~~~\vct{u}^{(0)} = \vct{0}, ~~~ \vct{\kappa}^{(0)} = \vct{0}, \quad \forall\,t\in[0,\infty).
\end{equation}

To explore the stability of such trivial equilibrium, we next consider the formal asymptotic expansion of the mechanical fields in the small parameter $\varepsilon$ and retain, for the linearization of the governing equations, only terms of
order one in that parameter. Hence, we write
\begin{equation}
    \vct{r} = \zero{\vct{r}} + \varepsilon \uno{\vct{r}} + o(\varepsilon), \qquad
    \vct{u} = \varepsilon \vct{u}^{(1)} + o(\varepsilon), \qquad \db{i} = \db{i}^{(0)} + \varepsilon \db{i}^{(1)} + o(\varepsilon),
\end{equation}
for the filament's configuration, for the strain vector, and for the directors, respectively. Since $\vct{u} = u_i \db{i}$, we observe that 
\begin{equation}
    \uno{\vct{u}}=\uno{u}_1\eb{x} + \uno{u}_2\eb{y} + \uno{u}_3\eb{z} ,
\end{equation}
whereas the linearization of equation~\eqref{eq:vectors_uw} gives
\begin{equation}
    \label{eq:lin_strain}
\partial_s \uno{\db{i}} = \uno{\vct{u}}\times  \zero{\db{i}} .
\end{equation}
By enforcing the constraint $\uno{\db{3}} = \partial_s \uno{\vct{r}}$ into~\eqref{eq:lin_strain} for $i=3$, we obtain
\begin{equation}
    \partial_s^2 \uno{\vct{r}}= \uno{\vct{u}}\times\eb{z} ,
\end{equation}
which provides
\begin{equation}
    \label{eq:lin_pos_strain}
\partial_s^2\uno{r}_1 = \uno{u}_2,\qquad\partial_s^2\uno{r}_2 = -\uno{u}_1,\qquad \partial_s^2\uno{r}_3 = 0 .
\end{equation}
Notice that the fields $\uno{r}_1$, $\uno{r}_2$ and $\uno{r}_3$ correspond to the Cartesian components of the incremental displacement, in light of the fact that $\db{1}^{(0)} = \eb{x}$, $\db{2}^{(0)} = \eb{y}$ and $\db{3}^{(0)} = \eb{z}$.
The linearized boundary conditions at the clamp~\eqref{eq:boundary_clamped} read
\begin{equation}
\label{eq:linClamp1}
    \uno{\vct{r}}(0,t) = \vct{0} , \qquad
    \uno{\db{1}}(0,t) = \vct{0} , \qquad
    \uno{\db{2}}(0,t) = \vct{0} ,
\end{equation}
for any $t \geq 0$.
Also, the inextensibility constraint $\Vert \partial_s \vct{r} \Vert = 1$ implies
\begin{equation}
    \partial_s\zero{\vct{r}} \cdot \partial_s\uno{\vct{r}} = \eb{z} \cdot \partial_s\uno{\vct{r}} = \partial_s \uno{r}_3 = 0 ,
\end{equation}
so that equation~\eqref{eq:lin_pos_strain}$_3$, together with the boundary condition~\eqref{eq:linClamp1}$_1$, gives $\uno{r}_3 = 0$.
Thus, the incremental motion is only in the transverse directions.

We next expand the internal contact force as $\vct{C}=\varepsilon\uno{\vct{C}}+o(\varepsilon)$, so that the linearization of equations~\eqref{eq:blm_components_nondim} gives
\begin{equation}
\label{eq:blm_components_nondim_lin}
    \left\{
    \begin{aligned}
    &\partial_s \uno C_1 - \xi \, \zeta_1 \, \partial_{t}\uno r_1 = 0 ,\\
    &\partial_s \uno C_2 - \xi \, \zeta_2 \, \partial_{t}\uno r_2 = 0 ,\\
    &\partial_s \uno C_3 = 0 .
    \end{aligned}
    \right.
\end{equation}
The free-end boundary condition $\uno{\vct{C}}(1,t) = \vct{0}$ and equation~\eqref{eq:blm_components_nondim_lin}$_3$ imply $\uno{C}_3 = 0$.
Similarly, by expanding the spontaneous strain as $\vct{\kappa}=\varepsilon\uno{\vct\kappa} + o(\varepsilon)$, the linearization of equations~\eqref{eq:bam_components_nondim} gives
\begin{equation}
    \label{eq:bam_components_nondim_lin}
    \left\{
    \begin{aligned}
    &\beta_1\partial_s (\uno u_1-\uno \kappa_1) - \uno C_2 = 0 ,\\
    &\beta_2 \partial_s (\uno u_2-\uno \kappa_2) + \uno C_1 = 0 ,\\
    &\beta_3\partial_s \uno u_3 = 0 .
    \end{aligned}
    \right.
\end{equation}
Equation~\eqref{eq:bam_components_nondim_lin}$_3$ and the free-end boundary condition $\uno{\vct{M}}(1,t) = \vct{0}$ yield $\uno{u_3} = 0$, \emph{i.e.}, the incremental motion does not involve twist of the filament.
Next, substitution of equations~\eqref{eq:bam_components_nondim_lin} and~\eqref{eq:lin_pos_strain} into equations~\eqref{eq:blm_components_nondim_lin} gives
\begin{equation}
    \label{eq:linear_disp_final}
    \left\{
        \begin{aligned}
            &\beta_2 (\partial_s^4 \uno r_1-\partial_s^2\uno \kappa_2) + \xi \, \zeta_1 \, \partial_t \uno r_1=0 ,\\
            &\beta_1 (\partial_s^4 \uno r_2+\partial_s^2\uno \kappa_1) + \xi \, \zeta_2 \, \partial_t \uno r_2=0.
        \end{aligned}
    \right.
\end{equation}

We now observe that equations~\eqref{eq:lin_strain} and~\eqref{eq:lin_pos_strain} provide
\begin{equation}
    \partial_s \uno{\db{2}}\cdot \eb{z} = \uno{u_1} = -\partial^2_s \uno r_2,\qquad
\partial_s \uno{\db{1}}\cdot \eb{z} = -\uno{u_2} = -\partial^2_s \uno r_1 .
\end{equation}
On accounting for the boundary conditions~\eqref{eq:linClamp1}$_2$ and~\eqref{eq:linClamp1}$_3$ in performing the spatial integration of these relations, we obtain
\begin{equation}
     \uno{\db{1}}\cdot \eb{z} =  -\partial_s \uno r_1 , \qquad \uno{\db{2}}\cdot \eb{z} =  -\partial_s \uno r_2 ,
\end{equation}
so that the linearization of equations~\eqref{eq:evol_nondim_alt2} reads
\begin{equation}
    \label{eq:evol_nondim_lin}
    \left\{
        \begin{aligned}
            &\partial_t \uno \kappa_1 + \uno \kappa_1 = \chi_1 \,\partial_s \uno r_2,\\
            &\partial_t \uno \kappa_2 + \eta^{-2} \uno\kappa_2 = - \eta^{-3} \, \chi_1 \, \partial_s \uno r_1,
        \end{aligned}
    \right.
\end{equation}
where the scalar parameter $\chi_1 = \vct{\chi}_1 \cdot \eb{z}$ admits both positive and negative values.

The boundary conditions at the clamp $s=0$ read
\begin{equation}
\left\{
\begin{aligned}
& \uno{r}_1(0,t) = \partial_s \uno{r}_1(0,t) = 0, \\
& \uno{r}_2(0,t) = \partial_s \uno{r}_2(0,t) = 0,
\end{aligned}
\right.
\label{eq:eigProblem_BC1}
\end{equation}
for every $t \geq 0$ and follow from the relations~\eqref{eq:linClamp1}$_1$ and $\partial_s\uno{\vct{r}}(0,t) = \uno{\db{3}}(0,t) = \vct{0}$.
Instead, the remaining free-end boundary conditions at $s=1$ read
\begin{equation}
\left\{
\begin{aligned}
& \partial_s^2 \uno{r}_2(1,t) + \uno{\kappa}_1(1,t) = \partial_s^3 \uno{r}_2(1,t) + \partial_s \uno{\kappa}_1(1,t) = 0 , \\
& \partial_s^2 \uno{r}_1(1,t) - \uno{\kappa}_2(1,t) = \partial_s^3 \uno{r}_1(1,t) - \partial_s \uno{\kappa}_2(1,t) = 0 , 
\end{aligned}
\right.
\label{eq:eigProblem_BC2}
\end{equation}
for every $t \geq 0$ and follow from the conditions $\uno{\vct{M}}(1,t) = \vct{0}$ and $\uno{\vct{C}}(1,t) = \vct{0}$.
It is interesting to notice that the equations governing the incremental motion of the filament in the $x\!-\!z$ plane and in the $y\!-\!z$ plane are decoupled.
Also, they do not exhibit a dependence on the Poisson ratio of the constituent material, as they are purely flexural.

To proceed, we now assume separation of variables, so that 
\[
\uno r_1(s,\,t) =\hat{r}_1(s) \,e^{\omega_1 t} , \quad
\uno\kappa_2(s,\,t)=\hat{\kappa}_2(s)\,e^{\omega_1 t} , \quad
\uno r_2(s,\,t) =\hat{r}_2(s)\,e^{\omega_2 t} , \quad
\uno\kappa_1(s,\,t)=\hat{\kappa}_1(s)\,e^{\omega_2 t} ,
\]
with $\omega_1,\,\omega_2\in\mathbb{C}$, and recast the system of linear equations given by~\eqref{eq:linear_disp_final} and~\eqref{eq:evol_nondim_lin} as two decoupled eigenvalue problems
\begin{equation}
\label{eq:eigProblem}
    \begin{bmatrix}
        - \dfrac{\beta_2}{\xi \zeta_1} \dfrac{\mathrm{d}^4}{\mathrm{d}s^4} \quad &  \dfrac{\beta_2}{\xi \zeta_1} \dfrac{\mathrm{d}^2}{\mathrm{d}s^2}  \\[1.5em]
        - \dfrac{\chi_1}{\eta^3} \dfrac{\mathrm{d}}{\mathrm{d}s}  &  - \dfrac{1}{\eta^2} 
    \end{bmatrix}
    \begin{bmatrix}
        \hat{r}_1 \\[0.5em]
        \hat{\kappa}_2
    \end{bmatrix}
    =
    \omega_1
     \begin{bmatrix}
        \hat{r}_1 \\[0.5em]
        \hat{\kappa}_2
    \end{bmatrix} ,
    \qquad
    \begin{bmatrix}
        - \dfrac{\beta_1}{\xi \zeta_2} \dfrac{\mathrm{d}^4}{\mathrm{d}s^4}  \quad &  - \dfrac{\beta_1}{\xi \zeta_2} \dfrac{\mathrm{d}^2}{\mathrm{d}s^2}  \\[1.5em]
        \chi_1 \dfrac{\mathrm{d}}{\mathrm{d}s}  &  - 1 
    \end{bmatrix}
    \begin{bmatrix}
        \hat{r}_2 \\[0.5em]
        \hat{\kappa}_1
    \end{bmatrix}
    =
    \omega_2
     \begin{bmatrix}
        \hat{r}_2 \\[0.5em]
        \hat{\kappa}_1
    \end{bmatrix} .
\end{equation}
\begin{figure}
    \centering
    \includegraphics[width=\textwidth]{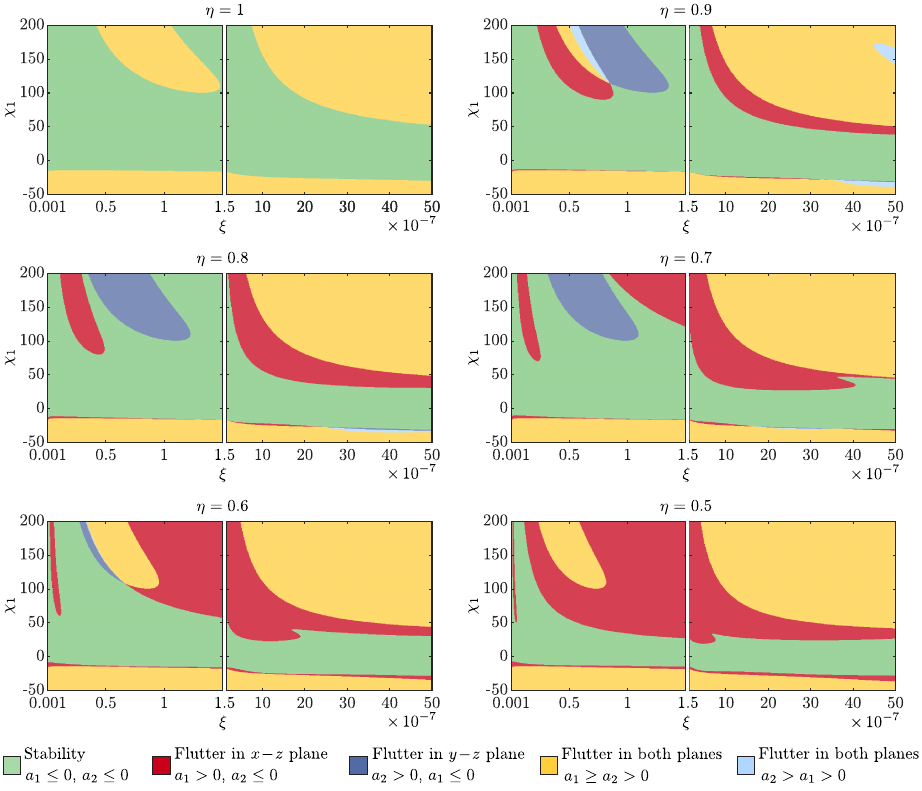}
    \caption{Stability diagrams in the $(\xi, \chi_1)$ plane, with $\xi \in [0.001, 50]\times 10^{-7}$ and $\chi_1 \in [-50, 200]$, for slenderness $\lambda = 50$ and cross-section aspect ratios $\eta = \{1, 0.9, 0.8, 0.7, 0.6, 0.5\}$. Five regions are identified and colour-coded to distinguish between stability (green), flutter in the $x\!-\!z$ plane (red), flutter in the $y\!-\!z$ plane (blue), and flutter in both planes, with oscillations developing faster in the $x\!-\!z$ plane (yellow) or in the $y\!-\!z$ plane (light-blue).}
    \label{fig:stab1}
\end{figure}
We resort to the finite element method to solve such eigenvalue problems, as detailed in Appendix~\ref{app:stability}, and focus here on presenting the relevant results.

Let $\omega_1^{\mathrm{max}} = a_1 \pm i \, b_1$ and $\omega_2^{\mathrm{max}} = a_2 \pm i \, b_2$ denote the eigenvalues with largest real part.
Varying the model parameters reveals distinct stability and instability regions: stability ($a_1 \leq 0$, $a_2 \leq 0$), flutter instability in the $x \! - \! z$ plane ($a_1 > 0$, $b_1 \neq 0$, $a_2 \leq 0$), flutter instability in the $y \! - \! z$ plane ($a_2 > 0$, $b_2 \neq 0$, $a_1 \leq 0$), flutter instability in both planes ($a_1 > 0$, $a_2 > 0$, $b_1 \neq 0$, $b_2 \neq 0$).
While stability is characterized by the decay of perturbation near equilibrium, flutter causes such perturbations to exponentially grow in amplitude, exhibiting oscillatory dynamics.
Far from equilibrium, these oscillations depart from the linear regime; thus, the behaviour of the system must be analysed through numerical experiments on the nonlinear model (see Section~\ref{EXP}).

We first investigate the stability of the mechanical system by fixing the slenderness $\lambda = 50$ and varying $\xi \in [0.001, 50] \times 10^{-7}$ and $\chi_1 \in [-50, 200]$ for $\eta = \{ 1, 0.9, 0.8, 0.7, 0.6, 0.5 \}$.
These parameter values were chosen to facilitate a natural comparison of the present results with those of the planar models discussed in~\cite{cicconofri_2023, boiardi_marchello_2024}.
Notice that values of $\eta > 1$ would lead to symmetric results, corresponding to exchanging the principal axes of the cross-section.
The results are shown in Fig.~\ref{fig:stab1}, where points of the parameter plane $(\xi,\chi_1)$ are colour-coded according to the stability of the system.
To adequately resolve the graphical features of the stability regions, we used different grid resolutions for $\xi \in [0.001, 1.5] \times 10^{-7}$ and $\xi \in [1.5, 50] \times 10^{-7}$.

For $\eta = 1$, the cross-section of the rod reduces to a circle, and problems~\eqref{eq:eigProblem}$_1$ and~\eqref{eq:eigProblem}$_2$ become identical.
Consequently, there is no difference between the motion in the $x\!-\!z$ plane and the motion in the $y\!-\!z$ plane; actually, if the system experiences flutter (yellow regions), oscillations can develop in any vertical plane.
We remark that $\chi_1 > 0$ and $\chi_1 < 0$ correspond to an applied electric field with $E_z > 0$ and $E_z < 0$, respectively. 
Clearly, flutter instability cannot occur if no external electric field is applied, \textit{i.e.}, for $\chi_1 = 0$.
By decreasing the cross-section aspect ratio, the instability regions for motions about the two principal cross-sectional axes separate, and the system exhibits a richer spectrum of behaviours. Here, flutter can occur exclusively in the $x\!-\!z$ plane (red regions) or in the $y\!-\!z$ plane (blue regions), or concurrently in both planes, with either $a_1 \ge a_2 > 0$ (yellow regions) or $a_2 > a_1 > 0$ (light-blue regions), such that oscillations develop faster in the $x\!-\!z$ or in the $y\!-\!z$ plane, respectively.
As $\eta$ decreases, the disparity in the flexural stiffnesses of the rod increases, thus favouring the motion in the $x\!-\!z$ plane.
This behaviour is reflected in the panels of Fig.~\ref{fig:stab1}, where the (red) regions corresponding to flutter in the weaker $x\!-\!z$ plane broaden as $\eta$ is reduced, whereas the (blue) regions corresponding to flutter in the stiffer $y\!-\!z$ plane are less affected.
Interestingly, the flutter regions depicted in Fig.~\ref{fig:stab1} for $\eta < 1$ exhibit self-similar behaviour.
Indeed, problem~\eqref{eq:eigProblem}$_2$ can be derived from problem~\eqref{eq:eigProblem}$_1$ via the following substitutions:
\begin{equation}
\label{eq:rescaling}
\hat{r}_1 \to \hat{r}_2 ,
\qquad
\hat{\kappa}_2 \to - \hat{\kappa}_1 , 
\qquad 
\chi_1 \to \eta \, \chi_1 , 
\qquad 
\xi \to \eta^2 \, \frac{\beta_2 \zeta_2}{\beta_1 \zeta_1} \, \xi ,
\qquad
\omega_1 \to \frac{1}{\eta^2} \, \omega_2 ,
\end{equation}
so that the response of the system in one plane can be determined from knowledge of its response in the other plane.

\begin{figure}
    \centering
    \includegraphics[width=\textwidth]{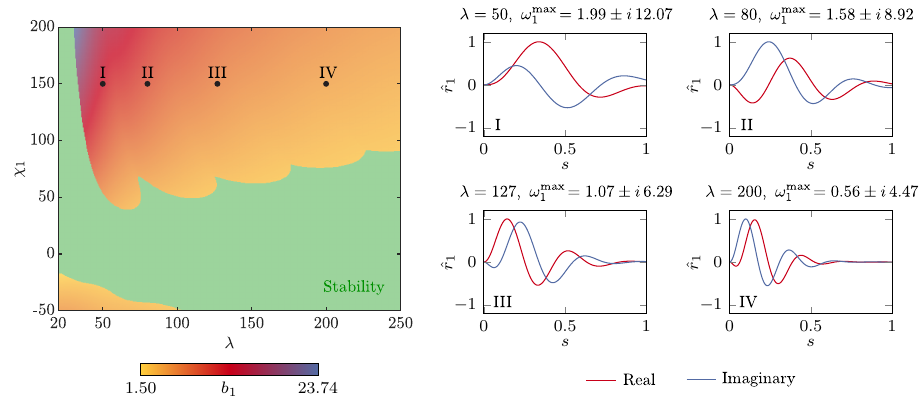}
    \caption{
    On the left, stability diagram in the $(\lambda, \chi_1)$ plane, with $\lambda \in [20, 250]$ and $\chi_1 \in [-50, 200]$, for $\xi = 50 \times 10^{-7}$ and cross-section aspect ratio $\eta = 1$. The stability region is depicted in green, while the flutter regions are colour-coded with the imaginary part, $b_1$, of the leading eigenvalue. Corresponding eigenfunctions are reported on the right for $\chi_1 = 150$ and $\lambda = \{50, 80, 127, 200\}$ to highlight the dependence of the flutter mode on the filament's slenderness.
    }
    \label{fig:stab2}
\end{figure}

Having established the influence of the cross-section aspect ratio on system stability, we fix $\eta = 1$ and $\xi = 50 \times 10^{-7}$ to explore the behaviour of the system as the slenderness parameter $\lambda$ varies.
Results for $\lambda \in [20, 250]$ and $\chi_1 \in [-50, 200]$ are reported in the left panel of Fig.~\ref{fig:stab2}, where the stability region is highlighted in green and the flutter regions are colour-mapped to the imaginary part of the leading eigenvalue.
It is worth recalling that here flutter may occur in all vertical planes, since $\eta = 1$.
We observe that the absolute value of $\chi_1$ for the onset of the instability generally increases with the slenderness.
The instability threshold exhibits a peculiar shape, suggesting the occurrence of a qualitative shift in the system behaviour at specific values of $\lambda$ for given $\chi_1$.
In support of this observation, we report in the right panel of Fig.~\ref{fig:stab2} the eigenfunctions $\hat{r}_1$ corresponding to $\omega_1^{\mathrm{max}}$ for $\lambda = \{ 50, 80, 127, 200\}$ and $\chi_1 = 150$.
As the slenderness increases, the number of inflection points in the leading mode increases, accumulating near the clamp ($s=0$), while the free-end ($s=1$) becomes increasingly flat.
Moreover, we highlight that real and imaginary parts of the eigenfunctions are not proportional, so that their modulation in time results in nonreciprocal periodic shape changes, which exponentially increase in amplitude in the flutter regime.

\section{Numerical experiments from the nonlinear model}
\label{EXP}

The linear analysis reported in the previous section describes small perturbations around the trivial equilibrium,  which can be explored using a semi-analytical approach. Conversely, the study of large deflections and complex motions far from equilibrium is only accessible through the numerical solution of the nonlinear governing equations. In this regime, the complexity of the system makes a systematic exploration of the parameter space unfeasible. Hence, we focus on representative test cases, which are selected to highlight the richness of the system's dynamics, such as the interplay between two-dimensional and three-dimensional motions and the role of nonlinearities in altering the stability thresholds predicted by the linear analysis. We remark that, consistently with the linear stability analysis presented above, all the numerical experiments discussed in the following sections are performed for an electric field aligned with the $z$-axis.

\subsection{Two-dimensional motions}

We begin this exploration with a filament of slenderness $\lambda=50$ and elliptical cross-section of aspect ratio $\eta= 0.9$, and set $\xi = 50 \times 10^{-7}$ and $\nu = 0.5$ to study the motion of the system for distinct values of the forcing parameter $\chi_1$. 
Although the results presented here were all obtained for $\nu = 0.5$, numerical experiments over the range $\nu \in [0.35,0.5]$ showed no qualitative differences. 

\begin{figure}[h]
    \centering
    \includegraphics[width=\textwidth]{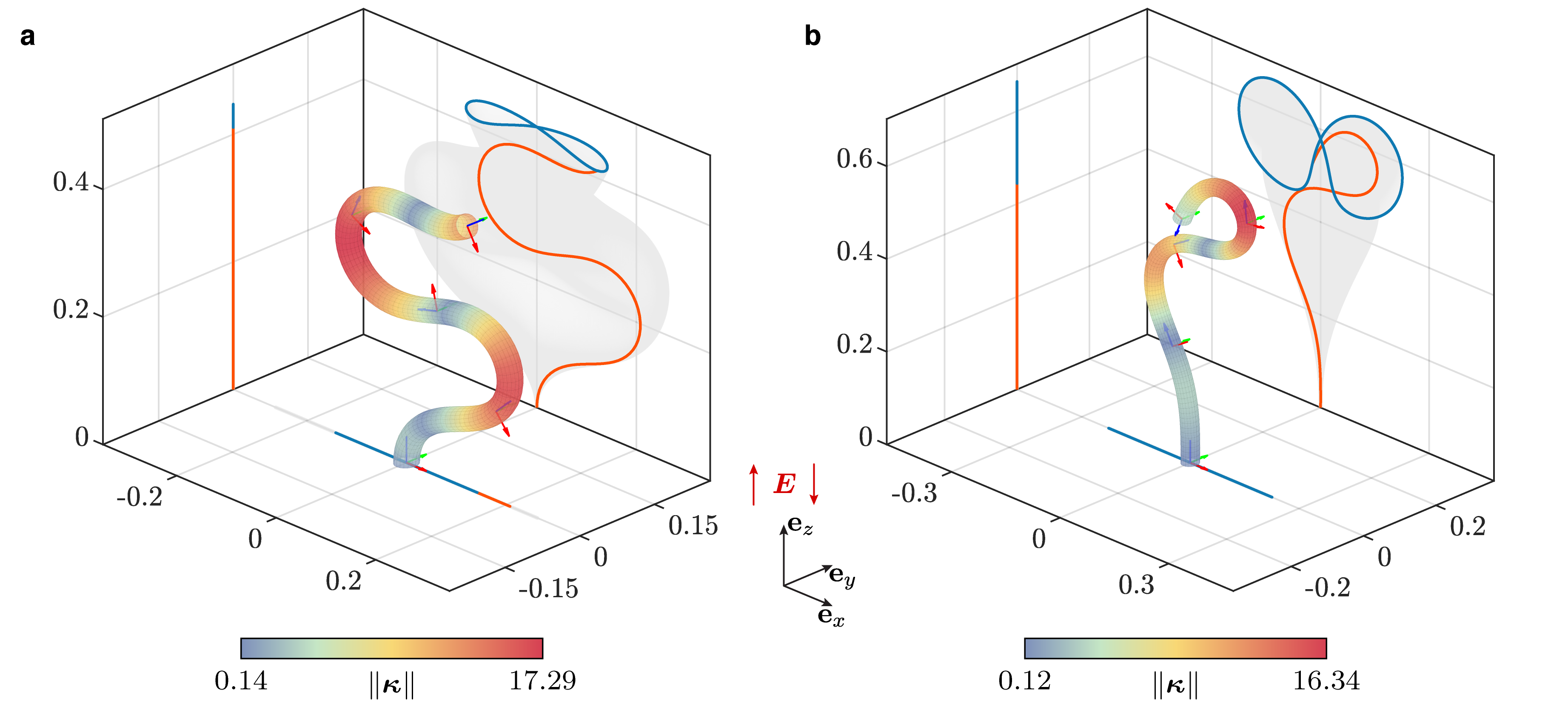}
    \caption{
        Two-dimensional nonlinear motion of a filament with slenderness $\lambda=50$, cross-sectional aspect ratio $\eta=0.9$, and $\xi = 50\times 10^{-7}$, for (a) $\chi_1 = 100$ and (b) $\chi_1 = -33$. The panels show the spatial configuration of the filament, endowed with the cross-sectional directors, its projections onto the Cartesian planes (orange curves), and the tip's trajectory (blue loops). The grey shaded regions represent the envelope of the motion, while the colour bars indicate the magnitude of the spontaneous curvature vector. 
    }
    \label{fig:planar_modes}
\end{figure}

Inspection of the relevant stability diagram in Fig.~\ref{fig:stab1} (top-right panel) reveals that the system is stable for $0 < \chi_1 \lessapprox 38$. Above this critical value, flutter instability occurs first in the $x\!-\!z$ plane and later also in the $y\!-\!z$ plane, as $\chi_1 \gtrapprox 51$. Under this condition a variety of flutter modes become feasible, and the perturbation that triggers the instability has a strong influence on the selected mode and the subsequent nonlinear dynamics. 
Perturbing the trivial equilibrium with a planar velocity field in the $x$ direction, we observe the emergence of planar motions in the $x\!-\!z$ plane, as reported in Fig.~\ref{fig:planar_modes}a for $\chi_1 = 100$ and in Fig.~\ref{fig:planar_modes}b for $\chi_1 = -33$. 
Similarly, a perturbation in the $y$ direction gives rise to planar flutter modes in the $y\!-\!z$ plane. 
These flutter modes reproduce in the three-dimensional setting the results already discussed in~\cite{cicconofri_2023,boiardi_marchello_2024}.
While a filament with circular cross-section can exhibit two-dimensional flutter modes on any vertical plane, this symmetry is broken for an elliptical cross-section, so that planar flutter is feasible only about the principal axes of the cross-section, \textit{i.e.}, in the $x\!-\!z$ and $y\!-\!z$ planes.

\begin{figure}
    \centering
    \includegraphics[width=\textwidth]{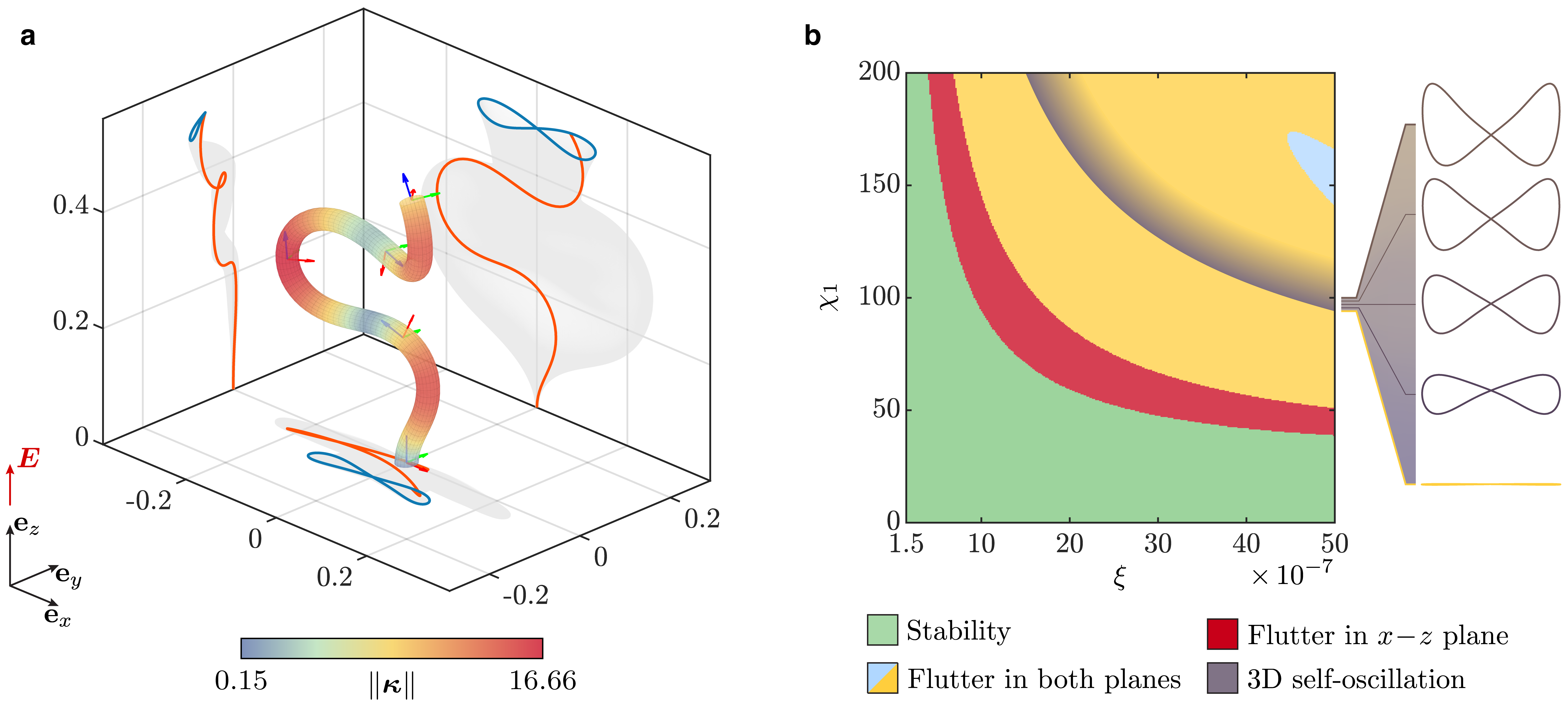}
    \caption{
        Three-dimensional nonlinear motion (a) of a filament with slenderness  $\lambda=50$, cross-section aspect ratio $\eta=0.9$, and $\xi = 50 \times 10^{-7}$ for $\chi_1 = 100$, and relevant stability diagram (b) showing the threshold for this flutter mode to occur (purple region) as computed from nonlinear numerical simulations. The panel also reports the projection onto the $x\!-\!y$ plane of the filament's tip trajectories for $\xi = 50 \times 10^{-7}$ and for $\chi_1 = \{100,\,98.5,\,97,\,95.5,\,94\}$.}
    \label{fig:3d_modes_1}
\end{figure}

\subsection{Three-dimensional motions}

Perturbing the trivial equilibrium of the system for the same parameters as above with a non-planar velocity field, we observe the emergence of a three-dimensional oscillatory motion, as shown in Fig.~\ref{fig:3d_modes_1}a. 
This mainly develops in the $x\!-\!z$ plane, consistently with the lower stability threshold for flutter in that plane, as compared to the other.
It is worth emphasizing that such nonlinear motion cannot be interpreted merely as a superposition of the planar flutter modes in the $x\!-\!z$ and $y\!-\!z$ planes, nor does its onset simply follow from the loss of stability by flutter in those planes.
Indeed, at early times, the system develops a three-dimensional motion consistent with the linear stability analysis. However, after a transient characterized by oscillations of growing amplitude, the leading flutter mode (that in the $x\!-\!z$ plane) prevails and the system tends toward a planar limit cycle.
\begin{figure}[h]
    \centering
    \includegraphics[width=\textwidth]{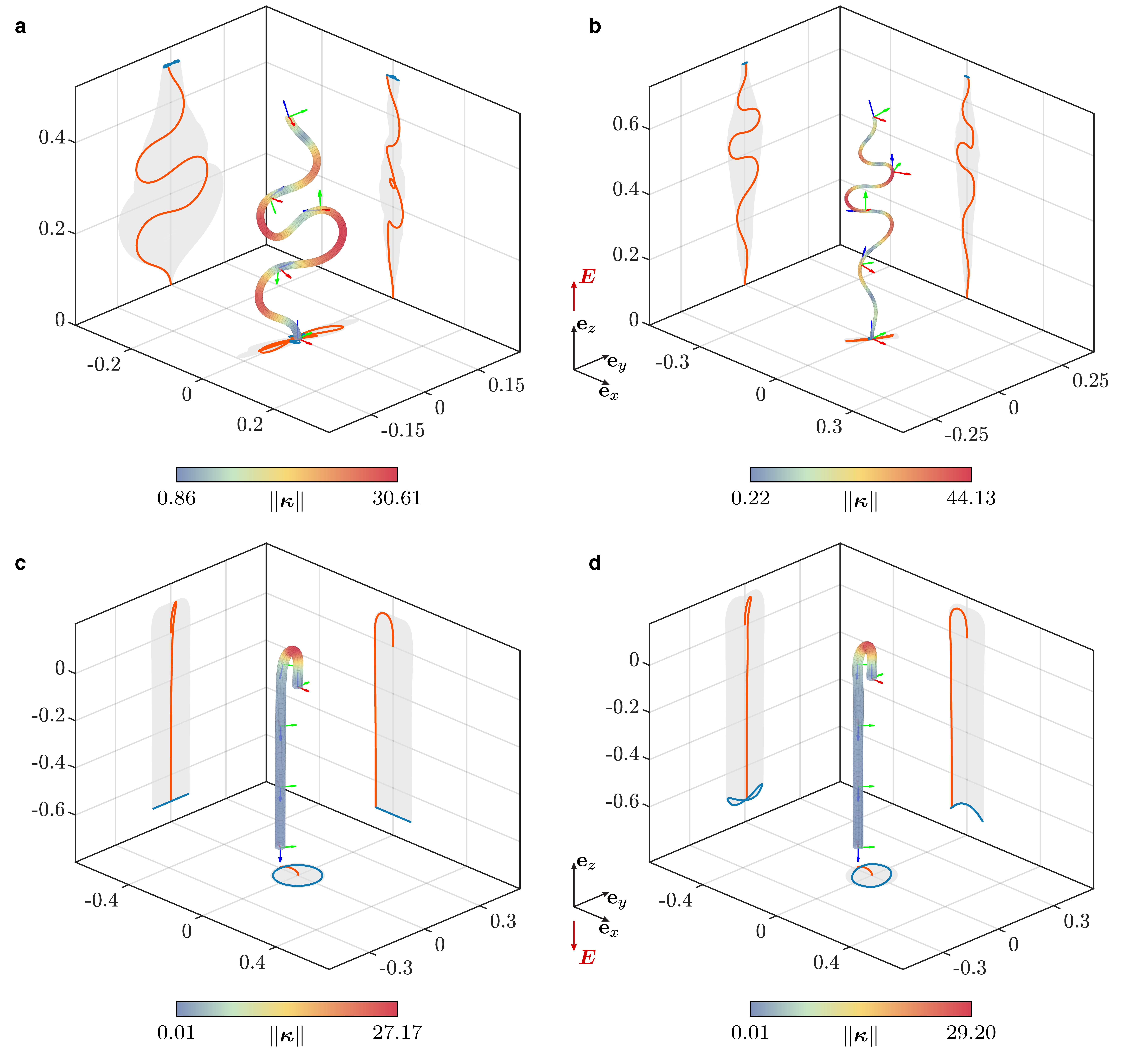}
    \caption{
        Three-dimensional nonlinear motions of a filament for $\xi = 50 \times 10^{-7}$ and for distinct values of $\lambda$, $\chi_1$, and $\eta$. The panels (a) and (b) allow us to compare two motions characterized by different inflection points as the filament's slenderness increases, obtained for $\eta=1$ and (a) for $\lambda = 127$ and $\chi_1=117$, and (b) for $\lambda = 200$ and  $\chi_1=160$. The panels  
        (c) and (d) show motions obtained for $\lambda=50$, with $\chi_1 =-33$ and for cross-section aspect ratio $\eta = 1$ in (c) and $\eta = 0.95$ in (d). In all panels, the filament is endowed with the cross-sectional directors,
        the grey shaded regions represent the envelope of the motion, while the colour bars indicate the magnitude of the spontaneous curvature vector.
    }
    \label{fig:3d_modes_2}
\end{figure}
The response of the system to an initial perturbation is substantially modified for values of $\chi_1$ sufficiently above the linear stability threshold for both planar modes. Under this condition, the numerical analyses reveal that the above planar limit cycle is not stable and that the system evolves to a persistent three-dimensional motion, see Supplementary Video~1.
This secondary bifurcation from large amplitude planar oscillations is strongly influenced by nonlinearities and cannot be predicted solely on the basis of the linear stability analysis.
Thus, the relevant stability threshold was determined through the numerical continuation of the nonlinear equations, varying the model parameters $\xi$ and $\chi_1$ and taking the fields corresponding to a three-dimensional motion as initial datum. Along the continuation, the disappearance of the three-dimensional motion was detected by a sharp transition in the aspect ratio of the projection onto the $x\!-\!y$ plane of the filament's tip trajectory, as shown on the right of Fig.~\ref{fig:3d_modes_1}b for the representative values of $\chi_1 = \{100,\,98.5,\,97,\,95.5,\,94\}$ and for $\xi = 50\times 10^{-7}$. 
The figure also reports the computed stability threshold and the (purple) region where the three-dimensional motion may occur.
As a point of comparison between two- and three-dimensional motions, it is instructive to evaluate the average power expended by the filament on the viscous fluid, that is
\begin{equation}
    \mathcal{\overline{P}} = - \frac{1}{T} \int_0^T\!\!\int_0^1 \vct{f}(\sigma,\tau) \cdot \vct{v}(\sigma,\tau)\,\,\mathrm{d}\sigma \,\mathrm{d}\tau ,  
\end{equation}
where $T$ denotes the (dimensionless) period of the considered motion.

Examining the representative cases of Fig.~\ref{fig:planar_modes}a and Fig.~\ref{fig:3d_modes_1}a, which were obtained for identical model parameters, this comparison reveals a decrease of about $7.6\%$ in the average power expenditure as the filament's motion transitions from two- to three-dimensional.

As the slenderness parameter $\lambda$ increases, more complex three-dimensional oscillatory motions emerge, characterized by an increased number of inflection points along the filament’s axis, see Fig.~\ref{fig:3d_modes_2}a and Fig.~\ref{fig:3d_modes_2}b, and Supplementary Video~2.
While this tendency is generally predicted by the linear stability analysis, the correspondence between the eigenfunctions reported in Fig.~\ref{fig:stab2} and the large deformations exhibited by the filament in the nonlinear regime is not strict.
In particular, a smooth evolution in the wavelength of the flutter eigenmodes is observed as $\lambda$ increases, but only some of these modes develop into three-dimensional, large amplitude motions, while others are suppressed or turn out to be unstable, so that the filament eventually settles into a different mode of deformation.
Also in this case, a non-planar perturbation induces a three-dimensional motion in the close-to-linear regime,  which transiently approaches a planar limit cycle. For suitable model parameters, however, this limit cycle is unstable, so that the system subsequently evolves into a persistent three-dimensional oscillatory motion.
In all numerical experiments, three-dimensional motions are consistently associated to a decrease of the average power expenditure, when compared to two-dimensional motions for the same model parameters. Specifically, this decrease is of about $4.2\%$ and $16.8\%$ for the cases reported in Fig.~\ref{fig:3d_modes_2}a and in Fig.~\ref{fig:3d_modes_2}b, respectively.

Despite the expected differences between the linear and the nonlinear regimes, the motions described so far retain the qualitative features of the underlying linear modes.
As the electric field is reversed, \textit{i.e.}, for $\chi_1 < 0$, the nonlinear motions originating from flutter substantially depart from the vertical configuration, thereby limiting the predictive capability of the linear analysis in this context.
In this respect, periodic two-dimensional motions can arise for negative values of the forcing parameter, as already reported in Fig.~\ref{fig:planar_modes}b. However, numerical simulations indicate that these oscillatory modes exist only over a limited range of parameters, and that the system may instead evolve toward alternative modes of deformation.
Indeed, for the set of parameters corresponding to the motion shown in Fig.~\ref{fig:planar_modes}b, a non-planar  perturbation brings the system towards a rotational motion characterized by hook-shaped configurations with curvature localized near the clamp, as shown in Fig.~\ref{fig:3d_modes_2}c and Fig.~\ref{fig:3d_modes_2}d. These configurations are not planar, especially where the curvature attains its maximum, and are such that the filament apparently rotates about the vertical axis while unwinding about its axis, by the evolution of the spontaneous curvatures, see Supplementary Video~3.
When the cross-section is circular, this motion is smooth and the filament's tip traces a perfect circle onto the $x\!-\!y$ plane, see Fig.~\ref{fig:3d_modes_2}c. When the cross-section is elliptical with moderate aspect ratio ($\eta = 0.95$), the motion is less symmetric and characterized by bumps in the trajectory of the filament's tip in the vertical direction, as shown in Fig.~\ref{fig:3d_modes_2}d.

\section{Conclusions and perspectives}
\label{CONC}

Building upon the planar formulation first introduced in~\cite{cicconofri_2023}, we develop a three-dimensional model for polyelectrolyte hydrogel filaments with elliptical cross-section subjected to electric fields.
Within the morphoelastic framework for Kirchhoff rods, the filament’s activity is encoded in spontaneous curvatures, which emerge at the macroscopic level from the migration of mobile ions, driven by the electro-chemical gradient, and the associated differential swelling in the cross‑sectional plane.
Guided by experimental observations, a phenomenological law is proposed for the evolution of the spontaneous curvatures, coupling the mechanical response of the filament to the applied electric stimulus. 
Aiming at applications of the proposed model to millimetre‑ and sub‑millimetre‑scale filaments in aqueous electrolyte solutions, we assume low Reynolds number hydrodynamics, so that fluid–structure interactions are captured through local resistive coefficients.

We examine the mechanical response of the filament to a constant electric field aligned with its axis. While the straight configuration is an equilibrium state under these conditions, the linear stability analysis of such equilibrium  reveals that the system undergoes flutter instability once the electric field exceeds a critical value. 
Due to the filament’s geometry, the first unstable mode consistently involves oscillations about the weak axis of the cross-section. 
As the electric field is further increased, flutter-induced oscillations also arise about the strong axis of the cross-section.
Near the equilibrium configuration, the interplay between these decoupled flutter modes gives rise to small-amplitude, non-planar, self-sustained oscillations of growing amplitude, which may ultimately develop into complex three-dimensional motions at large times.

To explore the post-critical regime, we solve numerically the nonlinear governing equations by the finite element method. 
Simulations show the emergence of three-dimensional motions in a subregion of the parameter space where three-dimensional flutter takes place. 
Depending on the model parameters, in particular the strength and direction of the applied electric field, and on the initial conditions, the filament exhibits either two-dimensional oscillations or persistent three-dimensional beating patterns and rotational motions.
Owing to their non-reciprocal nature, these periodic motions may be exploited for fluid transport or locomotion purposes. In this regard, for the test cases investigated in the present study, whenever two- and three-dimensional motions coexist the latter arise through a secondary bifurcation from the former. The occurrence of this secondary bifurcation is associated with a systematic decrease in the power expended by the filament on the viscous fluid, which apparently acts as an energetic criterion for the selection of the actual motion among the feasible ones. The variety of feasible motions is remarkable, especially in light of the time-independent nature of the external forcing. In this respect, the proposed flutter-based actuation strategy reduces control complexity, by effectively embodying it into the interaction of the system with its environment, thus opening new avenues in the design of hydrogel-based biomimetic systems.

This work represents a first step in extending state-of-the-art models for polyelectrolyte hydrogel filaments to three-dimensional slender active bodies. 
Future developments may involve the experimental investigation of this system, along with the exploration of three-dimensional flutter as a potential mechanism for propulsion and swimming.

\section*{Code availability statement}
All the COMSOL Multiphysics\textsuperscript{\textregistered}~6.3 implementation files and the custom post-processing MATLAB scripts used in the present study can be made available upon reasonable request.

\section*{Acknowledgements}
The authors acknowledge financial support from the Italian Ministry of University and Research (MUR) through the grant `Dipartimenti di Eccellenza 2023-2027 (Mathematics Area)'. G.N. acknowledges financial support from the European Union\,--\,Next Generation EU in the framework of the PRIN~2022 project no.~2022NNTZNM \lq Unveiling embodied intelligence in natural systems for bioinspired actuator design' (CUP:\,G53D23001240006). The authors are members of the \lq Gruppo Nazionale di Fisica Matematica' (GNFM) of the `Istituto Nazionale di Alta Matematica' (INdAM).

\section*{Description of Supplementary Videos}

All finite element simulations reported in the Supplementary Videos~1--3 were performed for $\nu = 0.5$ and $\xi = 50 \times 10^{-7}$. In all videos, $t$ denotes dimensionless time and the colour bars indicate the magnitude of the spontaneous curvature vector. In particular:
\begin{itemize}[topsep=7pt]
\setlength{\itemsep}{3pt}
    \item Supplementary Video~1 compares the two-dimensional (left) and the three-dimensional (right) periodic motion of a morphoelastic filament with $\lambda = 50$, $\eta = 0.9$, and $\chi_1 = 100$.
    \item Supplementary Video~2 compares the periodic three-dimensional motions of two morphoelastic filaments with $\eta = 1$, for $\lambda = 127$ and $\chi_1 = 117$ (left) and for $\lambda = 200$ and $\chi_1 = 160$ (right).
    \item Supplementary Video~3 compares the periodic three-dimensional motions of two morphoelastic filaments with $\lambda = 50$ and $\chi_1 = -33$, for $\eta = 1$ (left) and $\eta = 0.95$ (right). 
\end{itemize}

\begin{appendices}

\section{Numerical implementation of the linear stability analysis}
\label{app:stability}

The eigenvalue problems~\eqref{eq:eigProblem} were discretized
by the finite element method using the commercial software COMSOL Multiphysics\textsuperscript{\textregistered}~6.3. Accounting for the boundary conditions~\eqref{eq:eigProblem_BC1}-\eqref{eq:eigProblem_BC2} and denoting with a prime spatial differentiation,
the decoupled problems~\eqref{eq:eigProblem}$_1$ and~\eqref{eq:eigProblem}$_2$ were recast in weak form as
\begin{gather}
    \int_0^1 \left[ \beta_2 \, (\hat{r}_1''-\hat{\kappa}_2) \, \tilde{r}_1'' + \xi \, \zeta_1 \, \omega_1 \, \hat{r}_1 \, \tilde{r}_1 \right]  \mathrm{d}s = 0 , \qquad
    \int_0^1 \left( \omega_1 \, \hat{\kappa}_2 + \eta^{-2} \, \hat{\kappa}_2 + \eta^{-3} \, \chi_1 \, \hat{r}_1' \right)  \tilde{\kappa}_2 \, \mathrm{d}s = 0,
\end{gather}
and
\begin{gather}
    \int_0^1 \left[ \beta_1 \, (\hat{r}_2'' + \hat{\kappa}_1) \, \tilde{r}_2'' + \xi \, \zeta_2 \, \omega_2 \, \hat{r}_2 \, \tilde{r}_2 \right]  \mathrm{d}s = 0 , \qquad
    \int_0^1 \left( \omega_2 \, \hat{\kappa}_1 + \hat{\kappa}_1 - \chi_1 \, \hat{r}_2' \right) \tilde{\kappa}_1 \, \mathrm{d}s = 0,
\end{gather}
respectively, for suitable test functions $\tilde{r}_1 (s)$, $\tilde{r}_2 (s)$, $\tilde{\kappa}_1 (s)$ and  $\tilde{\kappa}_2 (s)$.
The discretization in space was performed using the Hermite finite element of order 3 for $\hat{r}_1$ and $\hat{r}_2$, ensuring $\mathcal{C}^1$-continuity of the lateral displacements, and the Lagrange finite element of order 2 for $\hat{\kappa}_1$ and $\hat{\kappa}_2$.
Denoting the continuous eigenfunctions as $\hat{\mathbf{X}} = (\hat{r}_1, \hat{\kappa}_2)^\mathsf{T}$ and $\hat{\mathbf{Y}} = (\hat{r}_2, \hat{\kappa}_1)^\mathsf{T}$, the discretized counterparts of~\eqref{eq:eigProblem} read
\begin{equation}
    \mathbb{K}_{N} \, \hat{\mathbf{X}}_{N} = \omega_1 \, \hat{\mathbf{X}}_{N} , \qquad
    \mathbb{H}_{N} \, \hat{\mathbf{Y}}_{N} = \omega_2 \, \hat{\mathbf{Y}}_{N} , 
\end{equation}
where $\hat{\mathbf{X}}_{N}$ and $\hat{\mathbf{Y}}_{N}$ are the vectors of nodal unknowns and $\mathbb{K}_{N}$ and $\mathbb{H}_{N}$ are the discrete stiffness matrices.
The domain $[0,1]$ was uniformly discretized in 50 mesh elements, leading to $N = 203$ degrees of freedom for each algebraic eigenvalue problem.
Once the parameters were fixed, the problems were solved using the LAPACK algorithm.
All discrete eigenvalues were computed and those with the largest real part,  $\omega_1^{\mathrm{max}}$ and $\omega_2^{\mathrm{max}}$, were identified along with their corresponding eigenfunctions.
In particular, the results in Fig.~\ref{fig:stab1} were obtained with a uniform parameter grid of $301 \times 501$ evaluations in the $(\xi, \chi_1)$ plane for $\chi_1 \in [-50, 200]$ and for both $\xi \in [0.001, 1.5]\times 10^{-7}$ and $\xi \in [1.5, 50]\times 10^{-7}$.
Likewise, the results in Fig.~\ref{fig:stab2} were obtained with a uniform parameter grid of $461 \times 501$ evaluations in the $(\lambda, \chi_1)$ plane for $\chi_1 \in [-50, 200]$ and for $\lambda \in [20, 250]$. The results from the finite element stability analyses were post-processed by custom MATLAB codes to produce Figs.~\ref{fig:stab1}\,--\,\ref{fig:stab2}.

\section{Numerical implementation of the nonlinear model}
\label{app:weak}

For the simulation of the nonlinear dynamics, the nonlinear governing equations presented in Section~\ref{MODEL:nondim} were discretized by the finite element method using the commercial software COMSOL Multiphysics\textsuperscript{\textregistered}~6.3. 
To this purpose, all governing equations were recast in weak form and discretized in space  as detailed in the following.

Equations~\eqref{eq:blm_components_nondim} from the balance of linear momentum were recast in weak form as
\begin{align}
    \int_0^1 \left( \partial_s C_1 - u_3 C_2 + u_2 C_3 - \xi\, \zeta_1 v_1 \right) \tilde{C}_1 \, \mathrm{d}s &= 0,\\[1mm]
    \int_0^1 \left( \partial_s C_2 + u_3 C_1 - u_1 C_3 - \xi\, \zeta_2 v_2 \right) \tilde{C}_2 \, \mathrm{d}s &= 0,\\[1mm]
    \int_0^1 \left( \partial_s C_3 - u_2 C_1 + u_1 C_2 - \xi\, \zeta_3 v_3 \right) \tilde{C}_3 \, \mathrm{d}s &= 0,
\end{align}
where $\tilde{C}_1(s)$, $\tilde{C}_2(s)$, and $\tilde{C}_3(s)$ are test functions for the internal contact force, as later specified.

As for equations~\eqref{eq:bam_components_nondim} from the balance of angular momentum, on accounting for the boundary condition~\eqref{eq:boundary_clamped}$_4$ and performing integration by parts, their weak form was obtained as
\begin{align}
     &\int_0^1 \left\{ \beta_1(u_1 - \kappa_1)\tilde{u}_1' + \left[\beta_2 u_3  (u_2-\kappa_2) - \beta_3 u_2  u_3 + C_2 \right] \tilde{u}_1 \right\} \, \mathrm{d}s +\beta_1[u_1(0)-\kappa_1(0)]\tilde{u}_1(0) = 0,\\[1mm]
    &\int_0^1 \left\{\beta_2(u_2 - \kappa_2)\tilde{u}_2' - \left[\beta_1 u_3 (u_1-\kappa_1) - \beta_3 u_1  u_3 + C_1 \right] \tilde{u}_2 \right\} \, \mathrm{d}s + \beta_2[u_2(0)-\kappa_2(0)]\tilde{u}_2(0) = 0 ,\\[1mm]
   & \int_0^1 \left\{ \beta_3 u_3 \tilde{u}_3' + \left[\beta_1 u_2 (u_1-\kappa_1) - \beta_2 u_1(  u_2 -\kappa_2) \right] \tilde{u}_3 \right\} \, \mathrm{d}s + \beta_3 u_3(0)\tilde{u}_3(0) = 0 ,
\end{align}
where $\tilde{u}_1(s)$ and $\tilde{u}_2(s)$ are test functions for the curvatures and $\tilde{u}_3(s)$ is the test function for the twist.

Finally, the weak form of equations~\eqref{eq:evol_nondim_alt} for the evolution of the spontaneous curvatures was obtained as
\begin{align}
     &\int_0^1 \left( \partial_t \kappa_1 + \kappa_1 + \vct{\chi}_1 \cdot \db{2} \right) \tilde{\kappa}_1 \, \mathrm{d}s = 0,\\[1mm]
     &\int_0^1 \left[ \partial_t \kappa_2 + \eta^{-2} \kappa_2 - \eta^{-3} (\vct{\chi}_1 \cdot \db{1}) \right] \tilde{\kappa}_2 \, \mathrm{d}s = 0 ,
\end{align}
where $\tilde{\kappa}_1(s)$ and $\tilde{\kappa}_2(s)$ are test functions for the spontaneous curvatures.

Since the Darboux strain vector $\vct{u}$ is determined by the balance of angular momentum, these equations are coupled with~\eqref{eq:vectors_uw} to determine the directors $\{\db{1},\db{2}\}$ in the cross-sectional plane and the tangent vector field $\db{3} = \db{1} \times \db{2}$, which in turn determines the filament's centreline $\vct{r}$ through the kinematic relation $\partial_s \vct{r} = \db{3}$. The relevant weak forms are not reported for brevity.

The spatial discretization was performed using Lagrange finite elements of order 1 for the spontaneous curvatures $\kappa_1$ and $\kappa_2$ and for the strain vector $\vct{u}$, Lagrange finite elements of order 2 for the local reference frame $\{\db{1},\db{2}\}$ and the internal contact force $\vct{C}$, and Lagrange finite elements of order 3 for the filament centreline $\vct r$. 
The domain $[0,1]$ was uniformly discretized in 30 mesh elements, leading to a total of 1008 degrees of freedom for the discretized system of equations. 
The mesh was refined to 60 elements for the simulation with $\lambda = 200$ in Fig.~\ref{fig:3d_modes_2}b to capture the more complex shapes of the filament.
Time integration was performed using a backward differentiation formula, with a time step of $1\times 10^{-3}$. The results from the nonlinear finite element analyses were post-processed by custom MATLAB codes to produce Figs.~\ref{fig:planar_modes}\,--\,\ref{fig:3d_modes_2} and the Supplementary Videos~1\,--\,3.

\end{appendices}

\bibliography{bibliography}

\end{document}